\documentclass[10pt, conference, compsocconf]{IEEEtran}
\usepackage{graphicx}
\usepackage{color,soul}
\let\labelindent\relax

\usepackage{enumitem}
\setlist[itemize]{leftmargin=*}

\graphicspath{ {./images/} }
\usepackage{cite}

\ifCLASSINFOpdf
\else
\fi
\usepackage[cmex10]{amsmath}
\usepackage{algorithm}
\usepackage{algorithmicx}
\usepackage[noend]{algpseudocode}
\begin{document}
%
\title{Incentive-Based Selection and Composition of IoT Energy Services}

\author{}

\author{\IEEEauthorblockN{Amani Abusafia, Athman Bouguettaya, and Sajib Mistry}
\IEEEauthorblockA{School of Computer Science\\
The University of Sydney, Australia\\
\{amani.abusafia, athman.bouguettaya, sajib.mistry\}@sydney.edu.au
}
}


%


\maketitle
\vspace{-4mm}

\begin{abstract}
We propose a novel \textit{incentive-based} framework for composing energy service requests. An incentive model is designed that considers the context of the providers and consumers to determine \textit{rewards} for sharing wireless energy. We propose a novel priority scheduling approach to compose energy service requests that maximizes the reward of the provider. A set of exhaustive experiments with a dataset and collected IoT users' behavior is conducted to evaluate the proposed approach. Experimental results prove the efficiency of the proposed approach.  

\end{abstract}

%
\IEEEpeerreviewmaketitle
\vspace{-5pt}
\section{Introduction}

\emph{Internet of Things (IoT)} is a paradigm where everyday objects, known as \emph{things}, are connected to the internet. These are usually equipped with sensors and actuators. A recent study has estimated that up to 75 billion connected IoT devices will be in use by 2025 \cite{NoofConnectedDevices}. Examples of IoT devices are smartphones, medical sensors, wearables, etc. IoT devices usually have augmented capabilities such as sensing, networking, and processing \cite{whitmore2015internet}. The ubiquity of IoT offers opportunities to crowdsource the capabilities of IoT devices \cite{ziegler2015internet}.

\emph{Crowdsourcing} IoT devices is the process of integrating  things by utilizing their data and functionality to create novel applications \cite{ziegler2015internet}. Crowdsourced applications include environmental monitoring, smart grids, and healthcare applications \cite{ziegler2015internet}. The abstraction of crowdsourced IoT devices may provide novel services \cite{bouguettaya2017service}. The \emph{IoT services} are defined by the \emph{functional} and \emph{non-functional} properties of IoT devices \cite{bouguettaya2017service}. Examples of \emph{IoT services} include Wi-Fi hotspot sharing and energy sharing\cite{lakhdari2018crowdsourcing} \cite{lakhdari2020composing}.

Energy sharing service, also known as \emph{Energy-as-a-Service (EaaS)}, is defined as transferring wireless energy among IoT devices using the service paradigm \cite{lakhdari2018crowdsourcing}. An \emph{energy provider} is a $thing$ that can share energy\footnote{We will use interchangeably the terms owner and provider to refer to the owner of the IoT device }. An \emph{energy consumer} is a $thing$ that requires energy. Consumers and providers are owned by users. Energy may be \emph{harvested} through {wearables} e.g. smart textile or smart shoes\cite{choi2017wearable}. The wearables may harvest energy from resources such as body heat or kinetic movement \cite{gorlatova2015movers}. For example, the PowerWalk kinetic energy harvester provides 10-12 watts on-the-move power\cite{gorlatova2015movers}. Users wearing a PowerWalk harvester on each leg can generate enough power to charge up to four smartphones from an hour walk at a comfortable pace \cite{PowerWalk}.The harvested energy could be shared with nearby IoT devices as EaaS with the newly developed technology called ``Over-the-Air wireless charging" \cite{OvertheAirCharger}. For example, Wattup technology, developed by Energous, enables wireless charging up to distances of 15 feet\footnote{https://www.energous.com/}. We focus on the use of wearables and IoT devices as energy providers. 

Providing EaaS has several advantages. For example, EaaS is a crowdsourced \emph{green} solution as it utilizes spare and renewable energy. Additionally, EaaS provides an alternative to using the power grid. Providing EaaS is a convenient solution compared to carrying power banks or plugging into a power outlet. {Recently, there has been an increased interest in the concept of} \emph{wireless crowdcharging} \cite{bulut2018crowdcharging}. There are several fields were EaaS is needed where a power outlet is not available. Examples include disaster management, entertainment, and emergency response \cite{PowerWalk}. The proposed environment typically consists of microcells, i.e. confined areas. A confined area may be any place where people aggregate such as coffee shops, restaurants, and movie theaters. In the environment, things may share energy using the Energy-as-a-Service model. We assume that a trust framework \cite{bahutair2019adaptive} has been implemented, hence, the IoT environment is secured for crowdsourcing EaaS.

As previously indicated, \emph{Energy-as-a-service model} is defined by its \emph{functional} and \emph{non-functional} (QoS) properties \cite{lakhdari2018crowdsourcing}. The function of the energy service is defined as the transfer of energy among  IoT devices. The non-functional properties describe the quality of service which would include energy capacity, location, and duration. To the best of our knowledge, existing research primarily focuses on the EaaS composition from a consumer perspective \cite{lakhdari2018crowdsourcing}.\emph{Our research focuses on the selection and composition of energy service requests from the provider's perspective}. 




One of the main challenges in composing EaaS is the resistance of providers. The resistance mostly occurs due to the limited resources of the IoT devices and lack of trust \cite{dhungana2020peer}. In many cases, providers may share energy \textit{altruistically} to help the environment. They can also be \textit{egoistic} since energy is a critical resource for their devices. \textit{Incentives} are a way to encourage providers to share their energy. Incentives additionally act as compensation for the providers' resource consumption. { To the best of our knowledge, there is no previous work done on incentivizing users to provide wireless energy sharing. Existing incentive models in a crowdsourced environment are not applicable in EaaS as the context of wireless energy sharing requires a customized model}\cite{dhungana2020peer}. \emph{We focus to design a novel incentive model in selecting and composing energy services requests}.\looseness=-1

We propose an incentive model that considers the characteristics that affect the willingness of the provider to participate. An incentive model relies on \emph{intrinsic or extrinsic} rewards to motivate participation. \emph{Intrinsic reward} is an internal satisfaction motivated by altruistic purposes. In contrast, \emph{extrinsic reward} is motivated by materialistic compensations \cite{muldoon2018survey}. In this work, we focus on computing the \emph{extrinsic} reward using an incentive model. \textit{Our incentive model will  not consider intrinsic rewards as they are self-motivated.} The incentive model will consider the \textit{contexts} of the providers and consumers that may affect the willingness of the provider such as the amount of requested energy and the required time to complete the energy service. \emph{Our aim is to compose the best set of energy requests using our incentive model}. \looseness=-1

\begin{figure}[!t]
\centering
\includegraphics[width=0.95\linewidth]{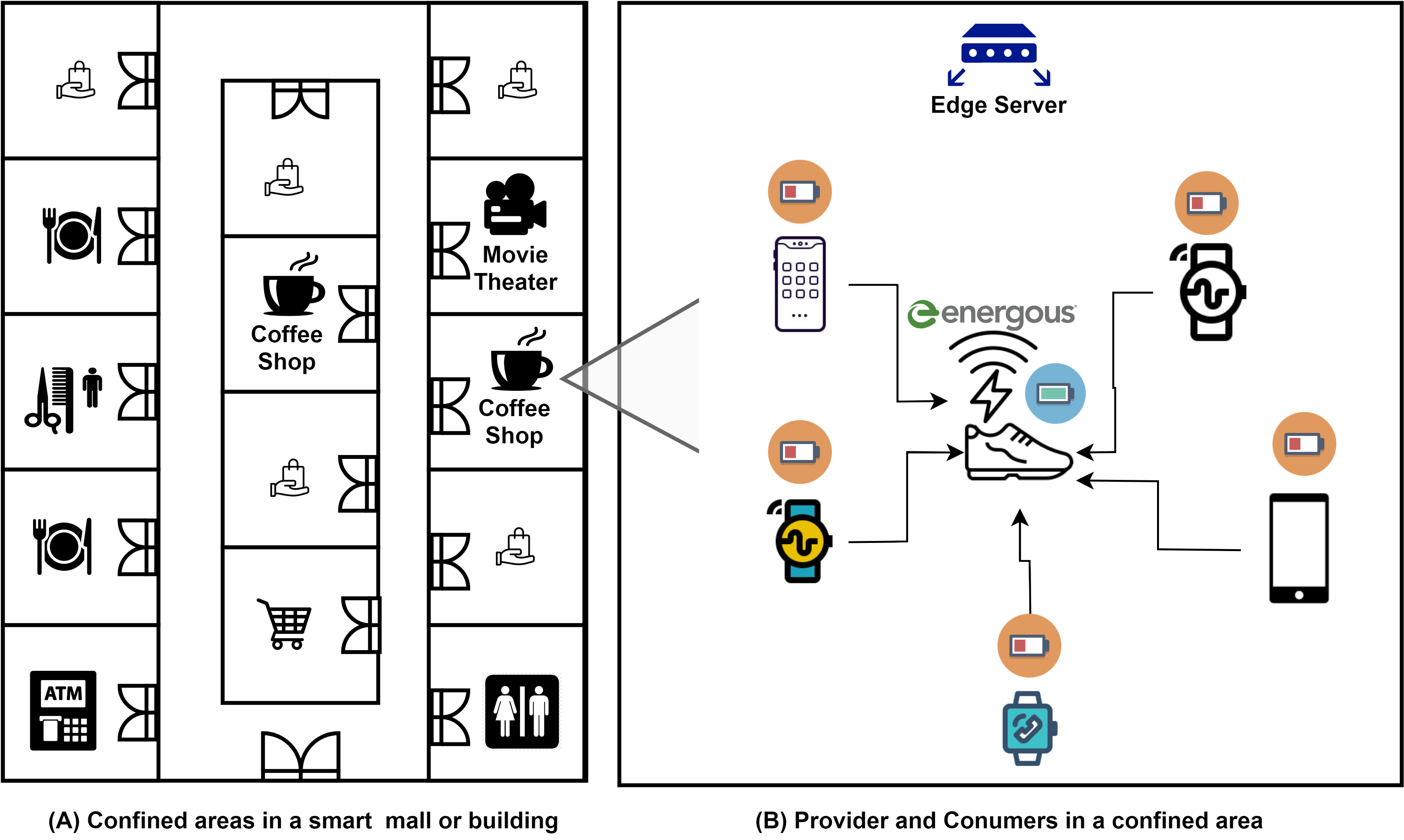}
\setlength{\abovecaptionskip}{0pt}
\setlength{\belowcaptionskip}{-2pt}
\caption{Wireless Energy Sharing Scenario}
\label{Scenariofig}
\end{figure}

\textbf{Motivation}. We consider an \emph{ incentive-based composition for energy service requests as a motivating scenario} (see Figure \ref{Scenariofig}). Assume that a geographic area, such as a building or a mall, is divided into $microcells$ where a microcell is a confined area such as a coffee shop or a restaurant (see Figure \ref{Scenariofig}a). Assume several IoT devices (providers and consumers) are distributed in a microcell (see Figure \ref{Scenariofig}b). The devices are assumed to be equipped with wireless energy transmitters and receivers such as Energous\footnote{https://www.energous.com/}. The distance between devices exchanging energy may reach the maximum distance (15 feet using Energous) to ensure successful wireless transmission. An owner of a device such as a smart shoe, which has spare energy,  would like to share the energy to available IoT devices. The provider receives requests from multiple energy consumers. All local energy requests and advertisements are processed at the edge i.e, a router associated with the microcell.  Consumers may have different requirements in terms of requested energy, charging speed, time availability, and location. The provider may resist offering their spare energy even if they had the intention because of multiple reasons. This may include the fear of needing excess energy at a later time. The provider may get motivated to give energy if an incentive is offered. Providers may be incentivized differently. For example, a provider may prefer providing the service in the shortest time. A second provider may desire to receive an extrinsic reward (monetary). Another provider may want to give all their energy to contribute to a green environment. \emph{We aim to select and compose energy service requests that will maximize the extrinsic reward of the provider. }

We focus on \emph{composing energy service requests to maximize the reward of the energy provider}. We assume a single energy provider may share their energy with multiple energy consumers within \emph{a specific time interval}. The \emph{composition} of energy service requests includes selecting the optimal requests that maximize the provider's reward. The reward of every energy request is calculated using an incentive model. Note that, we assume a \emph{static environment} where the provider and consumers do not move during the composition. To the best of our knowledge, there is no algorithm that addresses the problem of resistance in providers for participation while composing energy service requests\cite{dhungana2020peer}.

\noindent The main contributions of this paper are:
\begin{itemize}
    \item A Spatio-temporal selection of energy service requests.
    \item A scheduling inspired incentive-driven composition of energy service requests.
    \item A novel incentive model considering consumers' EaaS requests and providers' QoS that increases providers' participation.
\end{itemize}

\setlength{\textfloatsep}{0pt}
\vspace{-4pt}
\section{System Model and Definition}
We propose a formal model of our incentive-based energy service requests composition. {We consider the scenario of energy sharing in a microcell during a particular time interval} \emph{T}. We use the following definitions to formulate the problem.

\noindent
\textbf{Definition 1:  Energy-as-a-Service (EaaS).} We adopt the definition of EaaS in\cite{lakhdari2018crowdsourcing}.  An EaaS is defined as a tuple of $\{E_{ID}, E_{OwnerID}, F,Q\}$, where:
\begin{itemize}
    \item $E_{ID}$ is a unique energy service identifier
    \item $E_{OwnerID}$ is a unique owner identifier
    \item $F$ is the function of sharing energy by an IoT device owner $E_{OwnerID}$ via an IoT device $d$
    \item $Q$ is a tuple $\{P_{EC}, P_{Loc}, P_{ST}, P_{ET}\}$ where each attribute donates a $QoS$ property of $ES$ as following:
        \begin{itemize}
            \item $P_{EC}$ is the energy capacity the provider can share
            \item $P_{Loc}$ is the location of the provider $<x,y>$
            \item $P_{ST}$ is the start time of the provider stay in the microcell
            \item $P_{ET}$ is the end time of the provider stay in the microcell
        \end{itemize}
\end{itemize}
\noindent
\textbf{Definition 2: Energy Service Request (ER).} An ER request is a tuple of $\{ER_{ID}, ER_{OwnerID}, C_{BL}, C_{RE} , C_{ST}, C_{ET}, C_{Loc}\}$ where:
\begin{itemize}
    \item $ER_{ID}$ is a unique energy service request identifier
    \item $C_{OwnerID}$ is a unique consumer (owner) identifier
    \item $C_{BL}$ is the consumer's battery level at the time of sending the request
    \item $C_{RE}$ is the amount of requested energy by the consumer
    \item $C_{ST}$ is the start time of the consumer's stay in the microcell
    \item $C_{ET}$ is the end time of the consumer's stay in the microcell
    \item $C_{Loc}$ is the location of the consumer
\end{itemize}
\noindent
\textbf{Definition 3: Incentive.} The energy service provider receives an incentive reward $R$ by giving energy. Each incentive is earned as stored credit points. The incentive reward will be computed using an incentive model.

\begin{figure}[!t]
\centering
\includegraphics[width=0.8\linewidth]{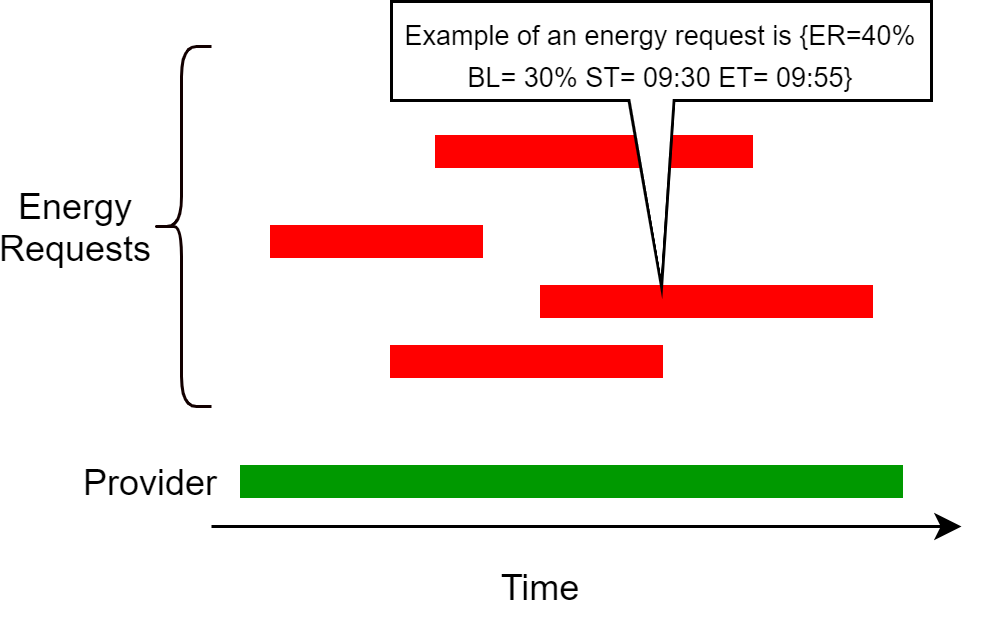}
\setlength{\abovecaptionskip}{-3pt}
\setlength{\belowcaptionskip}{5pt}
\caption{Example of Energy Requests and Service}
\label{EnergyRequests}
\end{figure}
\noindent
\textbf{Definition 4: Incentive-based energy service requests composition problem.}
We assume in a microcell, there exists an energy service $EaaS$ and a set of $n$ energy requests $ER = \{ER_{1},ER_{2},....,ER_{n}\}$ as shown in Figure \ref{EnergyRequests}. The $EaaS$ will be advertised by a provider $P$. The $EaaS$ will be represented using the aforementioned Definition 1. The energy requests are sent by consumers $C$. Each energy request (ER) is described using the aforementioned Definition 2. \emph{We formulate the composition of energy service requests into a service composition problem.} Composing energy service requests need to consider the spatio-temporal features of the service and requests. {Composing energy requests for a provider's} $EaaS$ requires the composition of energy requests $ER_{i} \in ER$ where $[C_{ST_{i}},C_{ET_{i}}] \subset [P_{ST},P_{ET}]$, $\sum C_{RE_{i}} \geq P_{EC}$ and the provider reward = $\sum R$ is the maximum. We use the following assumptions and definitions to formulate the problem.

\begin{itemize}
\item The IoT devices are equipped with wireless energy transmitters and receivers.
\item The composition considers the scenario of a single provider and multiple consumers.
\item The provider has fixed energy size during the composition.
\item The provider and consumers may have different time windows but consumers' time window must fall within the provider time window $T_c \in T_p$.
\item The provider and consumers have fixed locations $<x,y>$ for the whole duration of the energy service.
\item The provider can transfer energy to any consumer within their range (15 feet using Energous).
\item The provider transfers energy to one consumer at a time.
\item There is no energy waste while sharing. 
\end{itemize}
\begin{figure}[!t]
\centering
\includegraphics[width=0.7\linewidth]{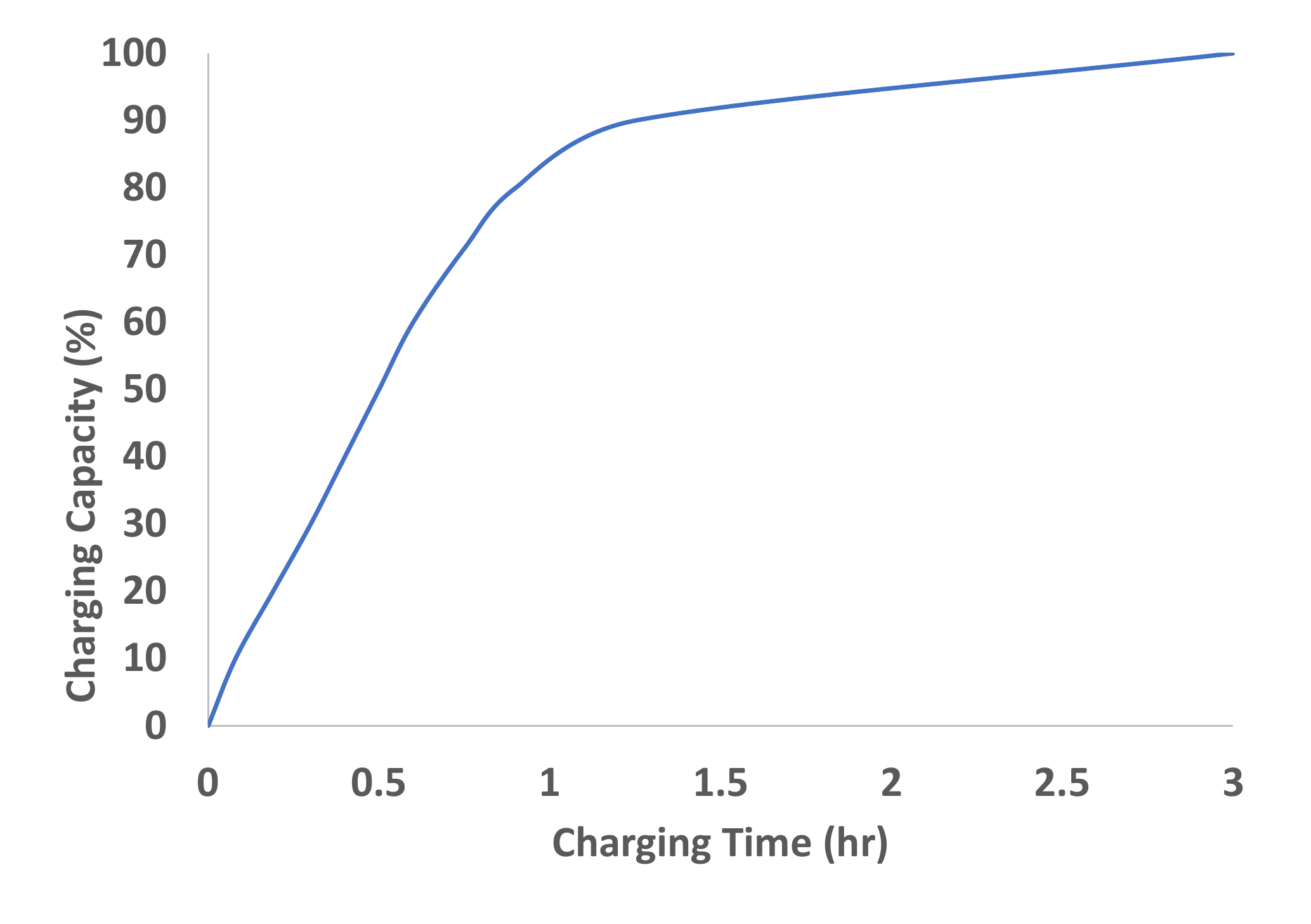}
\setlength{\abovecaptionskip}{-4pt}
\setlength{\belowcaptionskip}{6pt}

\caption{Charging time based on the capacity of lithium-ion 
batteries\cite{chargingSpeedImgRef}}
\label{chargingSpeed}
\end{figure}
\vspace{-4pt}
\section{Incentive Model}

\label{IM}
We define the incentive model to compute the reward $R$. The incentive model considers the attributes of the consumers and provider that may influence the resistance of the provider to give energy. The below attributes will be used to compute the reward.

\noindent\textbf{Attributes of Consumers' Context :}
\begin{itemize}
\item \textit{Battery level ($C_{BL}$)}: The battery level  affects the charging speed  as shown in Figure \ref{chargingSpeed}. A high charging speed encourages providers to give energy. The reward increases if the charging speed decreases. The charging speed is affected by the battery level of the consumer.  If the battery level is less than 80\% then the charging speed is high compared to battery level between 80\% and 99\%\cite{chargingSpeedImgRef}.  Additionally, consumers with low battery levels will be at risk of shutting down. The reward of giving energy to a consumer with a low battery should be more compared to a high battery level. We define the reward of the battery level ($Reward_{BL}$) as follow:  \textbf{if} $C_{BL} < 20\%$ \textbf{or} $C_{BL} > 80\% $ \textbf{then} $Reward_{BL}$  = 1  \textbf{ otherwise }  $Reward_{BL}$  = 0.5.
\item \textit{Requested Energy ($C_{RE}$)}: The amount of requested energy is considered in determining the reward. We only compute the reward of requested energy, if the provider can afford it. The reward of the Requested Energy ($Reward_{RE}$) is defined as follows: \textbf{if} $P_{EC} >= C_{RE}$ $ \textbf{then}$ $ Reward_{RE}  = C_{RE} / P_{EC}$ 
\item \textit{Stay-Time}: If a consumer is staying for a short time then they will be in a rush to get the requested energy. For example, if two consumers have the same energy request and similar attributes but differ in the availability time. The priority will be given for the one who is available for a shorter time. Therefore, the shorter the stay time of a consumer the higher the reward. We defined the reward of stay time as $Reward_{ST}  = |C_{time} - P_{time}| / P_{time} $ , where $P_{time}$ represents the total minutes of the provider time and $C_{time}$ represents the total minutes of the consumer time.
\end{itemize}

\begin{figure}[!t]
\centering
\includegraphics[width=0.8\linewidth]{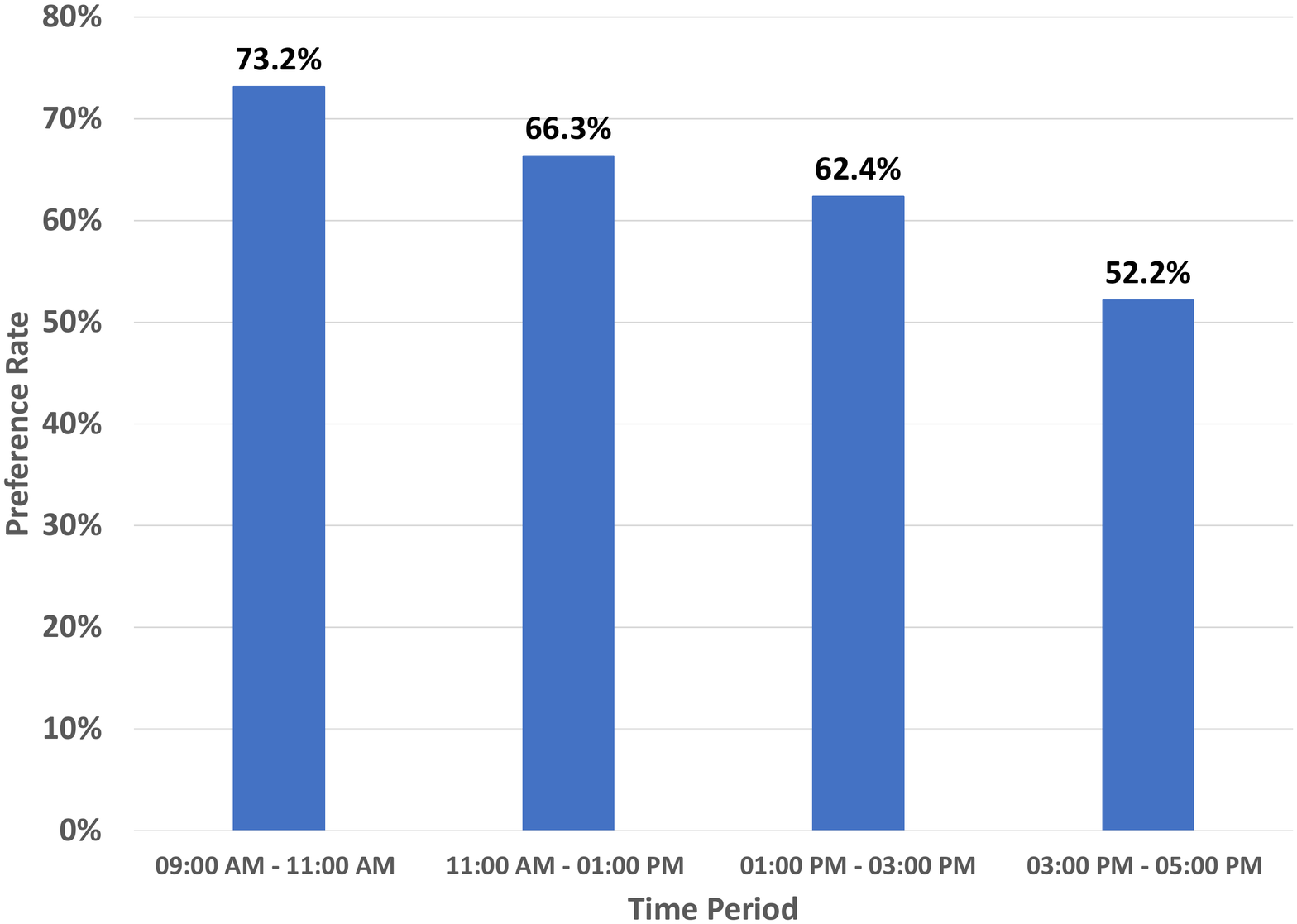}
\setlength{\abovecaptionskip}{-6pt}
\setlength{\belowcaptionskip}{-15pt}
\caption{The time period preferences for users to share energy.  }
\label{TPPreference}
\end{figure}
\begin{figure}[!t]
\centering
\includegraphics[width=0.8\linewidth]{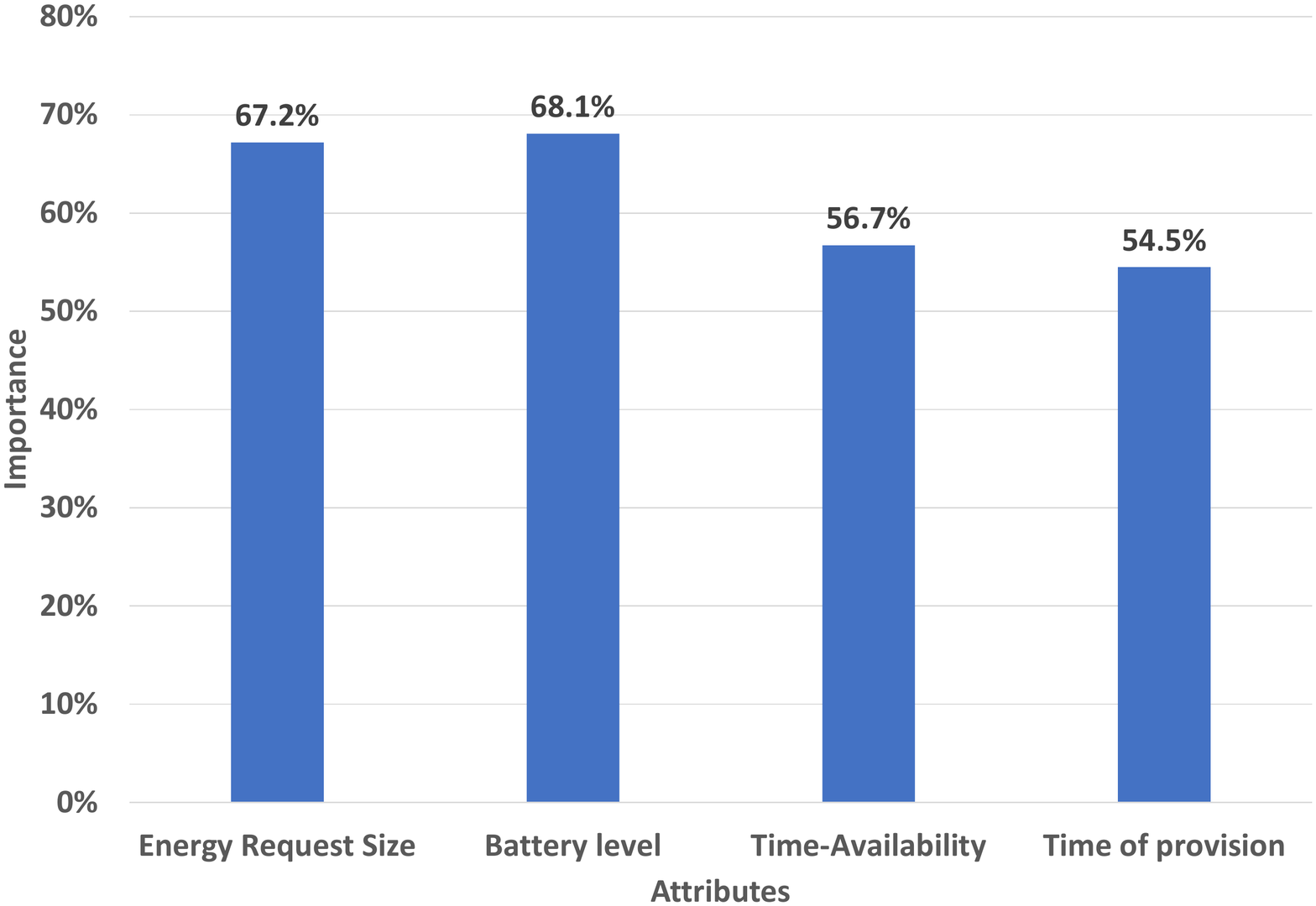}
\setlength{\abovecaptionskip}{-7pt}
\setlength{\belowcaptionskip}{4pt}
\caption{The average of the importance the incentive attributes based on preferences of users to share energy.  }
\label{RewardsWeight}
\end{figure}
\noindent\textbf{Attributes of Providers' Context }
\begin{itemize}
\item \textit{Time of Provision}: The time of providing energy influences the willingness of the providers. For example, the provider may be motivated to give the energy at the start of his stay more than at the end. This may occur due to having plenty of energy in the morning compared to the end of the day . We categorize the time of provision into four time periods $TP$ that is $TP_{1} = [09:00AM - 11:00AM],  TP_{2} = [11:00AM - 01:00PM], TP_{3} = [01:00PM - 03:00PM], TP_{4} = [03:00PM - 05:00PM]$. 

We use the crowdsourcing platform Amazon Mechanical Turk ($MTurk$) to compute the preferences of users in the time of provision\footnote{https://www.mturk.com/}. The use of MTurk is appropriate as incentives are human centric. Moreover, MTurk was used by other works as a platform to validate incentive models \cite{zhang2015incentives}.We designed a questionnaire that consists of set of scenarios. Each scenario starts by describing the energy sharing environment. We adopt an energy sharing environment where users can share energy among them through wireless technologies. MTurK workers are asked to consider themselves as energy service providers. We asked MTurk workers to select their preferred time period to accept energy requests from 09:00 AM to 05:00 PM. Figure \ref{TPPreference} represents the users' time period preference for sharing energy. The x-axis represents the defined time periods. The y-axis shows the percentage of users who selected a certain time period. For example, the figure shows that 73.2\% of participants preferred the time period from 09:00 Am to 11:00 Am. We normalized the time preference statistics to compute the reward of the Time of Provision ($Reward_{TP}$) as follow: We compute the resistance rate of each time period by subtracting the participation rate from 100, i. e. the resistance rate of $TP_{1}$ = 100\% - 73.2\% = 26.8\% . Then we normalized the resistance rate as the weight of the reward for $TP_{1}$. The normalization is computed by dividing the resistance rate of $TP_{1}$ by the total of resistance rates of all $TPs$.  We followed the same steps to calculate the reward of the Time of Provision ($Reward_{TP}$) which were as follow:\textbf{if} $TP_{i} \in TP_{1}$  \textbf{then} $Reward_{TP}$ = 0.18 \textbf{if} $TP_{i} \in TP_{2}$  \textbf{then} $Reward_{TP}$ = 0.23 \textbf{if} $TP_{i} \in TP_{3}$  \textbf{then} $Reward_{TP}$ = 0.26 \textbf{if} $TP_{i} \in TP_{4}$  \textbf{then} $Reward_{TP}$ = 0.21
\end{itemize}
\textbf{Energy Request Total Reward:} The {total reward} of an energy service request ($RewardER$) is the summation of all the rewards as follows:
\begin{equation}
\setlength{\abovedisplayskip}{3pt}
\setlength{\belowdisplayskip}{3pt}
 RewardER = \sum_{i \in n}^{}w_{i} \times Reward_{i} 
\end{equation}
 Where $n =\{BL,RE,ST,TP\} $and $w$ is the weight of each reward. $w$ is computed using MTurk as previously indicated. We asked users to scale the importance of each attribute when they accept energy requests. The scale value is between 0\% and 100\%, 0\% indicates an unimportant attribute while 100\% indicates a highly important attribute. Figure \ref{RewardsWeight} represents the average of the importance of each attribute. Then, we normalized the importance of the attributes to define the weight of each of them. We computed the normalization by dividing the importance of an attribute over the total of the importance of all the attributes. Therefore, the weight of each reward $w$ is \{0.27, 0.28, 0.23, 0.22\} in the same order of set $n$.  

\textbf{Provider Total Reward:}
\setlength{\abovedisplayskip}{6pt}
\setlength{\belowdisplayskip}{6pt}
The total reward of the provider is the summation of the  total reward of the selected  energy service requests as follows:
\begin{equation}
\label{ProviderRewardEq}
Provider Reward = \sum_{j=1}^{m} RewardER_{j} \end{equation}
Where $m$ is the number of selected energy requests $ER$ that will receive the energy from the provider. The energy requests will be selected using our composition framework.
\vspace{-15pt}
\section{Composition Framework for Energy Service Requests }

\begin{figure}[!t]
\centering
\includegraphics[width=0.8\linewidth]{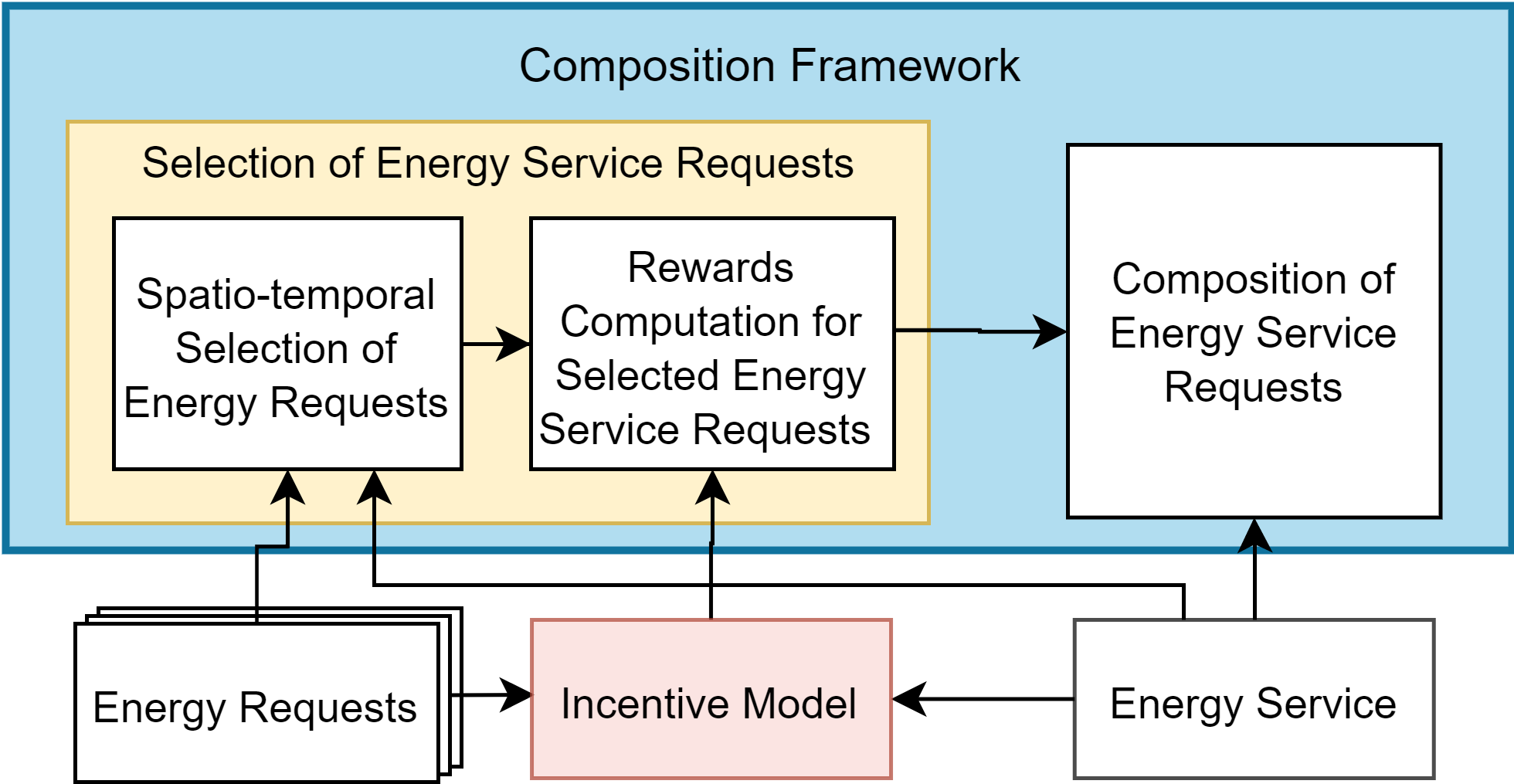}
\setlength{\abovecaptionskip}{2pt}
\setlength{\belowcaptionskip}{2pt}
\caption{The composition framework for energy requests.  }
\label{EnergyFrameWork}
\end{figure}

The framework of composing energy service requests will involve two phases: the selection of energy service requests and the composition of energy service requests (See Figure \ref{EnergyFrameWork}). The first phase consists of selecting the energy service requests that are composable with the energy service. Then calculating the reward for each of the selected energy service requests using the incentive model.  The second phase is composing the energy service requests that maximize the reward for the provider. Our composition algorithm is inspired by Priority-based scheduling algorithms\cite{kruse2007data}.

\subsection{Selection of Energy Service Requests}

The selection phase aims to select all the energy requests that are composable with the energy service of the provider.  The selection phase consists of two stages: Spatio-temporal selection of energy requests and rewards computation for the selected energy requests (See Figure \ref{EnergyFrameWork}). The first stage aims to select nearby energy service requests that can be served by the provider. The selection is accomplished using the spatial and temporal features of EaaS. Spatial composability service is required to allow the wireless energy transfer to occur between the consumers and the provider. An energy request is composable temporally if its duration falls within the time window of the energy provider. The last step in the first stage is to check if the requested energy can be provided by the EaaS. The second stage focuses on computing the reward for each selected energy service request. We compute the reward using the incentive model discussed in Section \ref{IM}.

Phase 1 in Algorithm \ref{alg:CER} describes the selection phase of energy service requests (Line 1-7). For every energy service request, the algorithm checks if the time window of the energy service request falls in the time interval of the energy service (Line 1-2). Line 3 calculates the distance between the energy request and the energy service. Line 4 checks if the distance between the energy request and energy service is below the maximum needed distance to transfer energy (15 feet using Energous). Line 5 checks if the energy service capacity is enough to provide the amount of requested energy. Line 6 calculates the reward value of the energy request using the previously defined incentive model (Section \ref{IM}). Line 7 adds the energy request with its reward value to the set of nearby energy service requests. 
\vspace{-2pt}
\subsection{{Incentive-based} Composition of Energy Service Requests.}
The composition of the energy service requests phase aims to compose the energy requests that incentivize the provider to share their energy. This is accomplished by selecting energy requests that maximize the reward of the provider. Phase 2 of Algorithm \ref{alg:CER} describes the composition of energy service requests (Line 8-18). Line 8 sorts the nearby energy requests in ascending order based on the start time then in descending order based on the reward. Lines 12-14 check for each request if the available energy of the provider can serve the requested energy, then it sums up the ER's reward as part of the provider's reward using equation (\ref{ProviderRewardEq}). Line 15 updates the available energy of the provider by subtracting the energy request $C_{RE}$ from the Provider energy $P_{EC}$. Line 16 updates the time of the provider so that no energy requests overlap as the provider can only share energy with one consumer.  Line 17 adds the request to the set of composed energy request.

\begin{algorithm}[t!]
\small
    \renewcommand{\algorithmicrequire}{\textbf{Input:}}
    \renewcommand{\algorithmicensure}{\textbf{Output:}}
    \caption{Incentive-Based Composition of Energy Service Requests}
    \label{alg:CER}
    \begin{algorithmic}[1]
        \Require
        $EaaS, ER$
        \Ensure $ER_{composition} , Reward$ 
        \Statex \textbf{Phase 1: Selection of Energy Service Requests}
        \For{$ER_{i} \in ER$}
            \If{$C._{ST}\geq P_{ST}$ \textbf{and} $C._{ET}\leq P_{ET}$}
          
                \State $ distance = \sqrt{(C_{loc.x}-P_{loc.x})^2 + (C_{loc.y}-P_{loc.y})^2} $
                         
                \If{$distance \leq MaxEnergyDistance$ }
                    \If{$C_{RE}\leq P_{EC}$}
                        \State $ER_{i.Reward} = Calculate\_Reward(er_{i})$ 
                        \State $NearbyER$.add($ER_{i} , ER_{i.Reward}$)
                    \EndIf 
                \EndIf 
            \EndIf
        \EndFor
        \Statex \textbf{Phase 2: Composition of Energy Service Requests}
        \State $NearbyER_{sorted} = sort(NearbyER, starting_time:ascending, Reward:descending) $
        \State $Reward = 0$
        \State $Provider_{ST} = P_{ST}$
        \State $Provider_{EC} = P_{EC}$
        \For{$er_{i} \in NearbyER_{sorted}$}
            \If{$C._{ST}\geq Provider_{ST}$}
                \State $Reward = Reward + er_{i.Reward}$
                \State $Provider_{ST} = C._{ET}$
                \State $Provider_{EC} = C_{RE}$
                \State $ER_{composition}.add(er_{i})$
            \EndIf
        \EndFor
        \State \Return $ER_{composition} , Reward$ 
    \end{algorithmic}
\end{algorithm}
\section{Experimental Results and Discussions}
\label{experiments}
We conduct two sets of experiments to test the proposed composition framework and incentive model.  The first set of experiments evaluates the performance of the proposed Incentive-Based composition algorithm. We describe the execution time and energy utilization of the proposed composition against other approaches.The second set of experiments evaluate the relevance of the incentive model's attributes on users' resistance to share energy. {Moreover, the experiments show the effect of using rewards on increasing participation in sharing energy.}

\subsection{Evaluation of the Composition Framework}
We evaluate the performance of the proposed Incentive-Based composition algorithm (IB) against the Brute Force approach (BF) and the First Come First Served (FCFS). We then conduct two sets of experiments to evaluate the execution time and energy utilization of each algorithm. 

The dataset used in the experiments of this section is published by IBM for a coffee shop chain with three branches in New York city\footnote{https://ibm.co/2O7IvxJ}. The dataset consists of transaction records of customers purchases in each coffee shop for April. We took from each record the transaction date, time, location, and coffee shop ID. We ran the experiments on the data of one of the branches. The dataset of each coffee shop consists of 16500 transaction records on average in the whole month. The number of transaction records each day is 560 on average in each coffee shop. We used the records as energy service requests and randomly generated the battery level and requested energy $C_{RE}$ for each request $C_{BL}$ between 1\% to 100\%.  We then generated 2000 energy service advertisements with random energy capacity $P_{EC}$, location $P_{Loc}$, and time $P_{ST,ET}$. Table \ref{ExpVariables} summarizes the experimental variables. 

\begin{table}[!t]
\centering
\caption{Experiments Variables}
\label{ExpVariables}
\renewcommand{\arraystretch}{1}
\begin{tabular}{l|l}
\hline
Variables & Value   \\ \hline
Total Energy Requests for coffee shop 1 in April            & 16830   \\ 
Energy Services                  &{2000}             \\ 
Duration of Services             & {10 - 200 minutes} \\ 
Duration of Energy Requests      & {5 - 30 minutes}   \\
Provided Energy                  & {50 - 100 \%}      \\
Requested Energy                 & {1 - 100 \%}       \\
Battery Level                    & {1 - 80\%}         \\  \hline
\end{tabular}
\end{table}

\begin{figure}[!t]
\centering
\includegraphics[width=0.8\linewidth]{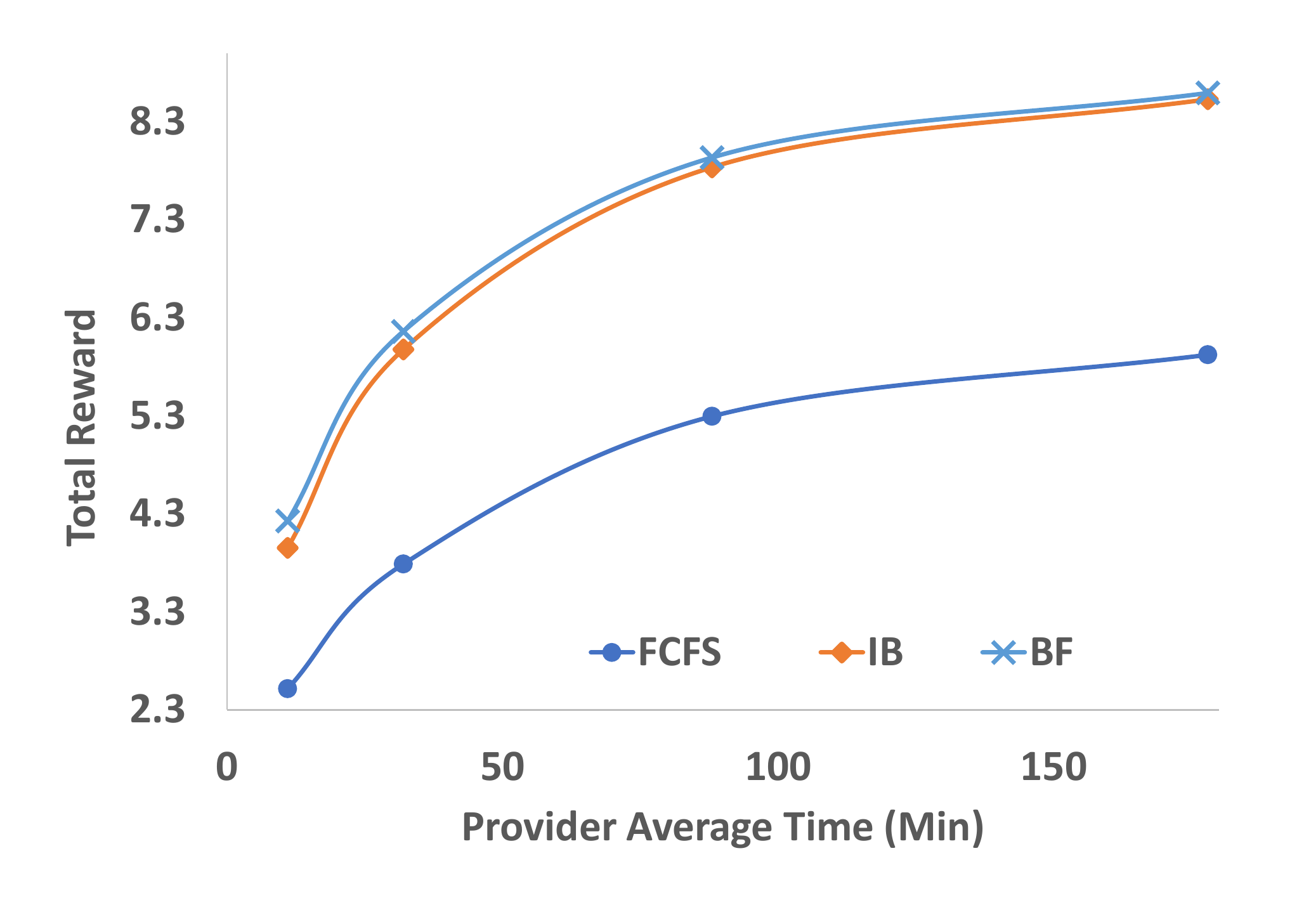}
\setlength{\abovecaptionskip}{-10pt}
\setlength{\belowcaptionskip}{3pt}
\caption{The average of Total reward for the Incentive-Based (IB), Brute Force (BF), and First Come First Served (FCFS) approaches.  }
\label{RewardAll}
\end{figure}

\begin{figure}[!t]
\centering
\includegraphics[width=0.8\linewidth]{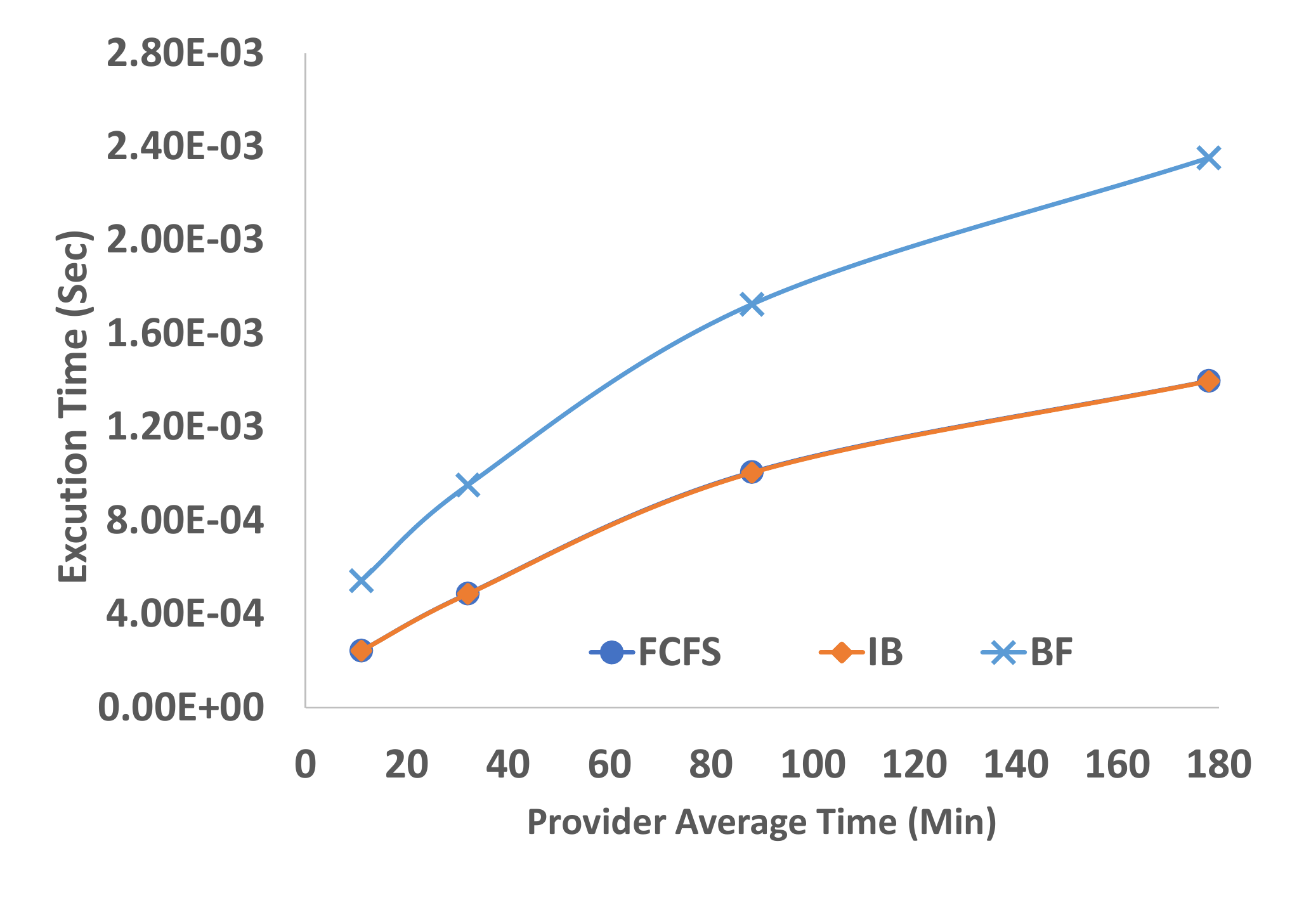}
\setlength{\abovecaptionskip}{-6pt}
\setlength{\belowcaptionskip}{-3pt}

\caption{The average of execution time for the Incentive-Based (IB), Brute Force (BF), and First Come First Served (FCFS) approaches.}
\label{ExeTime}
\end{figure}

In the first experiment, we compare the average total reward of the proposed Incentive-Based composition algorithms (IB) against the Brute Force approach (BF) and First Come First Served (FCFS)\cite{kruse2007data}.  We consider the Brute Force approach as the baseline. For the Brute Force approach, we retrieve all the possible compositions of the energy service requests. Then, we select the composition that has the maximum total reward. In the case of FCFS, the approach selects the energy requests based on their start time regardless of their reward value\cite{kruse2007data}. Fig. \ref{RewardAll} presents the average total reward for each algorithm. The x-axis in Fig. \ref{RewardAll} {represents the average of the provider's staying time}. The proposed algorithm (IB) performs better than the FCFS in terms of total reward. Our proposed algorithm maximizes on average the rewards by 22\% compared to the FCFS as it considers the reward value of each energy request. The Brute Force (BF) as a baseline gives the best result but the proposed algorithm gives close results and the BF comes with a cost of higher execution time as shown in Fig. \ref{ExeTime}.

\begin{figure}[!t]
\centering
\includegraphics[width=0.8\linewidth]{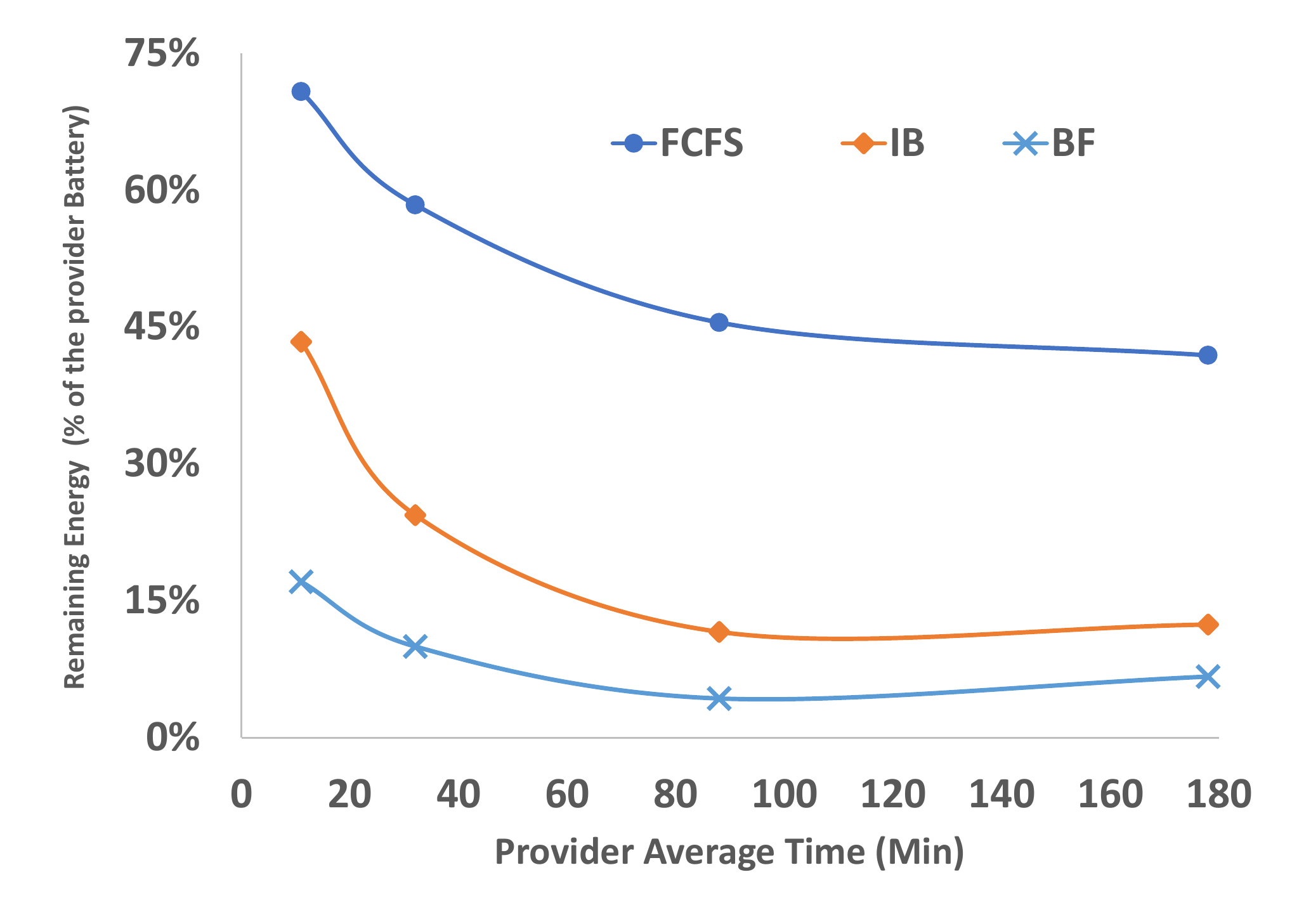}
\setlength{\abovecaptionskip}{-6pt}
\setlength{\belowcaptionskip}{-5pt}

\caption{The average of providers' remaining energy for the Incentive-Based (IB), Brute Force (BF), and First Come First Served (FCFS) approaches.}
\label{RemainEnergy}
\end{figure}

In the second experiment, we compare the average of the remaining energy of the providers after composition using the proposed Incentive-Based composition algorithms (IB) against the Brute Force approach (BF) and the First Come First Served (FCFS). The experiment test which approach has better energy utilization.  Fig. \ref{RemainEnergy} shows the average remaining energy of the providers for each algorithm. The x-axis in figure \ref{RemainEnergy} {represents the average of the provider's staying time}. The graph shows the proposed algorithm (IB) performs better than the FCFS in terms of utilizing the energy of the provider. Our proposed algorithm minimizes on average the remaining energy by 31\% as it considers the reward value of each energy request. As previously mentioned, the reward value takes into consideration the size of the energy request.  Additionally, the Brute Force (BF) as a baseline gives the best result but BF comes with a higher execution time as shown in Fig. \ref{ExeTime}.

\vspace{-5pt}
\subsection{Evaluation of the Incentive Model}
We present a set of experiments to evaluate the effectiveness of the incentive model. As indicated in section \ref{IM}, We use MTurk to validate the impact of the attributes on the participation of users in sharing energy. Moreover, MTurk is used to show the effect of using rewards on increasing participation in sharing energy. We designed a questionnaire that consists of a set of scenarios. These scenarios are presented to MTurk workers to test each attribute such as energy request size and time of provision. Each scenario starts by describing the energy sharing environment. We adopt an energy sharing environment where users can share energy among them through wireless means. MTurK workers are asked to consider themselves as energy service providers. Moreover, workers are asked to accept or reject requests by varying the attributes to measure the resistance of each attribute.\looseness=-1

All the described experiments in this subsection were designed with the following structure: Given a certain scenario, if a worker rejects a request, then we ask them if an extra reward would make them accept that request. If a worker still rejects the request we ask them if the amount of reward would change their decision. A total of 175000 questionnaires were answered by workers on MTurk. All the experiments in this subsection compare the participation rate of users in sharing energy without an incentive (No reward), with an incentive (Reward) and with deciding the amount of incentive (Amount of reward).  

\begin{figure}[!t]
\centering
\includegraphics[width=0.8\linewidth]{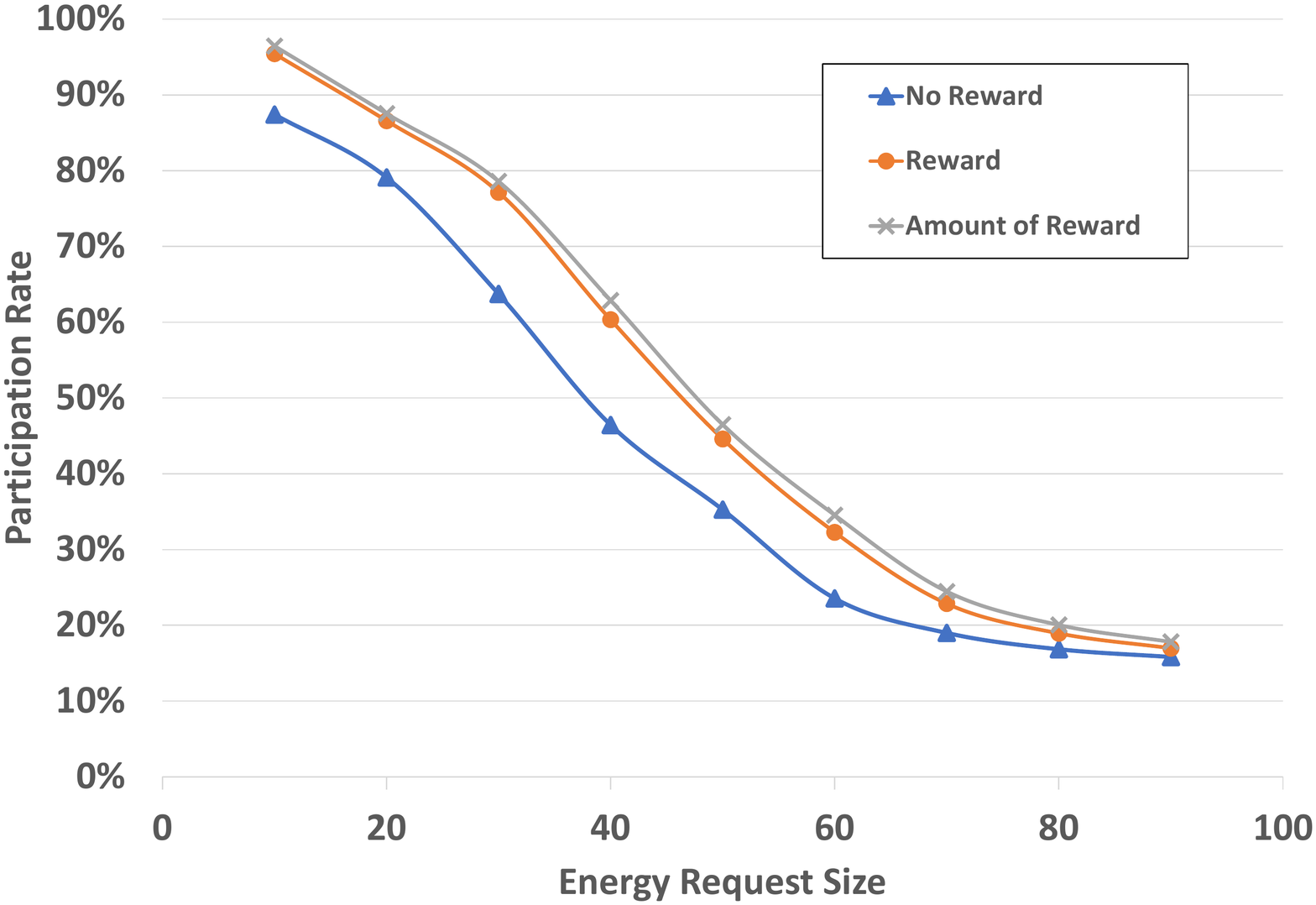}
\setlength{\abovecaptionskip}{-6pt}
\setlength{\belowcaptionskip}{-10pt}
\caption{{The participation rate of users in sharing energy with and without a reward.}}
\label{EnergyRequestSize}
\end{figure}

In the first experiment, we asked Mturk workers to accept or reject energy requests with different energy request size. The energy request size varied from 10\% to 90\% of the provider battery by increasing 10\% for each question. Figure \ref{EnergyRequestSize} presents the participation rate of users in sharing energy with different requests. The x-axis represents the energy requests' size. The y-axis shows the percentage of users who accepted the energy request. The figure reflects the effect of the energy request size attribute on the willingness to share energy. Additionally, the figure shows that providing a reward increases the participation of users in sharing energy.

\begin{figure}[!t]

\centering
\includegraphics[width=0.8\linewidth]{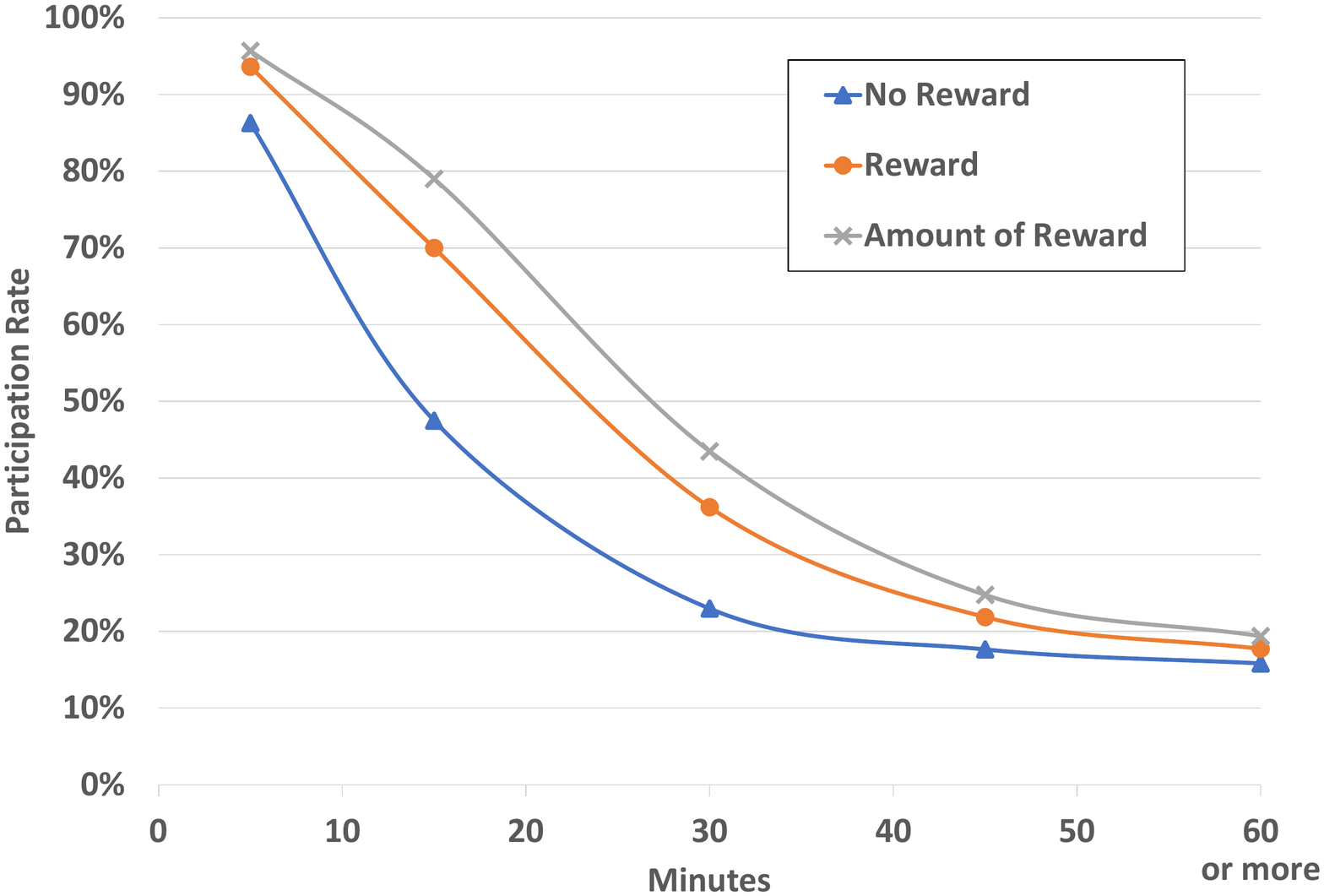}
\setlength{\abovecaptionskip}{-12pt}
\setlength{\belowcaptionskip}{-10pt}
\caption{{The participation rate of users in sharing energy with and without a reward.}}

\label{chargingTime}
\end{figure}

In the second experiment, we asked Mturk workers to accept or reject energy requests with different charging times. The charging time is used to test the effect of the battery level attribute. As previously indicated, the battery level affects the speed of charging and therefore it affects the time to charge a device.  The energy request charging time varied between 5 to $\geq$60 minutes. Figure \ref{chargingTime} presents the participation rate of users in sharing energy with different requests. The x-axis represents the energy requests charging time. The y-axis shows the percentage of users who accepted the energy request. The figure reflects the effect of the charging time on sharing energy. Additionally, the figure shows that giving a reward increased the participation of users in sharing energy.

In the third experiment, we asked Mturk workers to assume they are in a coffee shop and willing to share energy. Additionally, we asked them to select their preferred time period to share energy. If a time period was not selected, we ask them if an extra reward would make them accept the rejected time period. If a worker still rejects the request we ask them if the amount of reward would make them change their mind.  Figure \ref{chargingTime} presents the participation rate of users in sharing energy with different requests. The x-axis represents the time period to provide energy for a request. The y-axis shows the percentage of users who accepted the energy request. The figure shows that giving a reward increased the participation of users in sharing energy.

\begin{figure}[!t]
\centering

\includegraphics[width=0.8\linewidth]{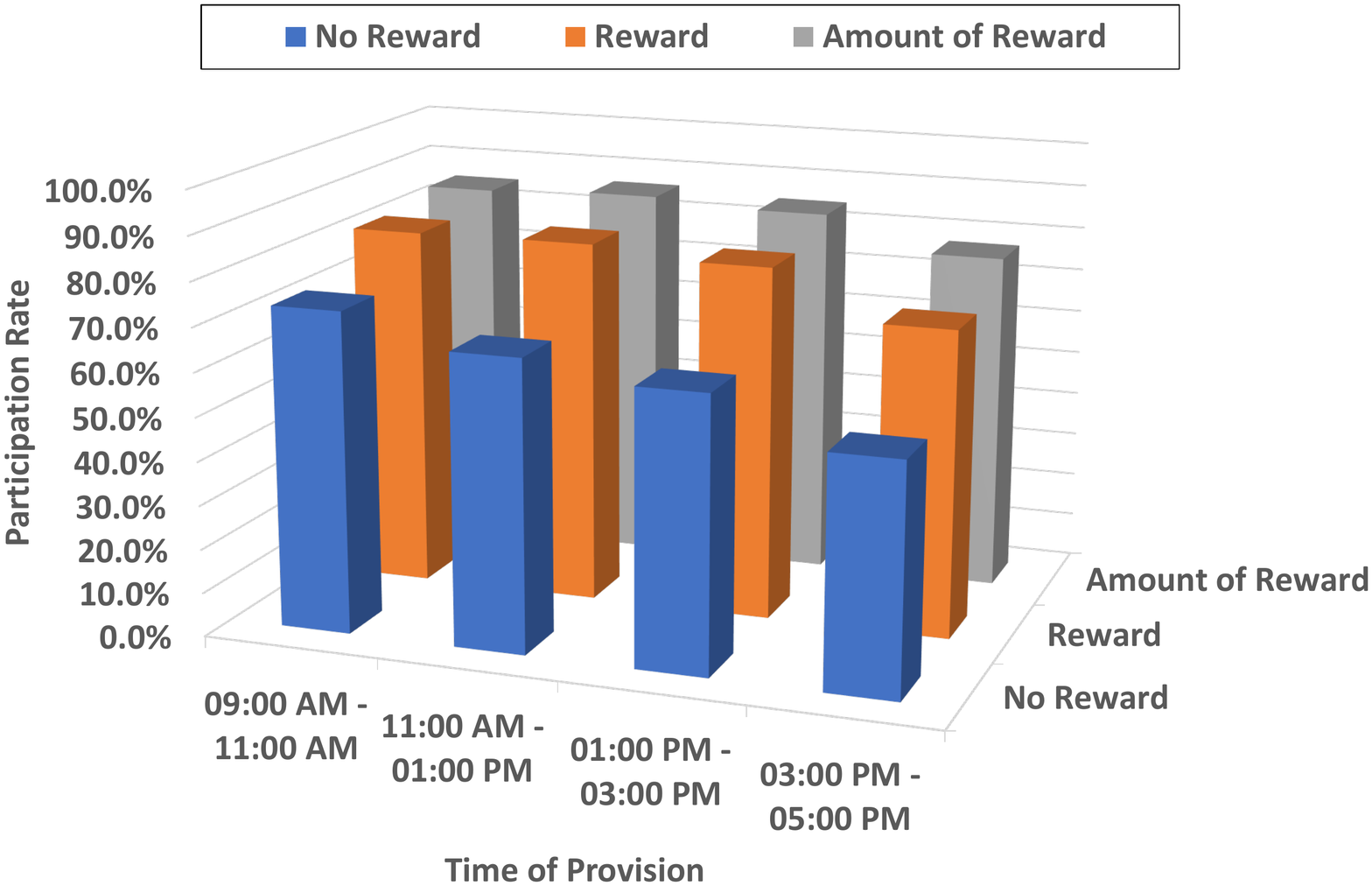}
\setlength{\abovecaptionskip}{-14pt}
\setlength{\belowcaptionskip}{2pt}
\caption{{The participation rate of users in sharing energy with and without a reward.}}
\label{TPParticipation}
\end{figure}

\vspace{-5pt}
\section{Related work}
\vspace{-5pt}
There are recent studies on energy sharing among mobile devices\cite{dhungana2020peer}.
Energy Sharing was used for several objectives including optimal energy usage, balancing energy distribution and energy sharing for content delivery \cite{dhungana2020peer}.  Optimal energy usage aims to reduce charging from outlets by chagrining from other devices' energy \cite{dhungana2019exploiting}. Balancing energy distribution was achieved in a mobile network by exchanging energy between devices \cite{nikoletseas2017wireless} or by constructing a star network structure \cite{madhja2018peer}. Energy sharing is also used in content delivery where nodes pass by massages to a destination node \cite{dhungana2019energy}. The massage is transferred with energy to motivate the nodes to carry the content to the destination node.  Most of the existing research assumed that mobile devices are motivated to give energy \cite{dhungana2020peer}. composing energy services requests while considering incentives is yet to be address\cite{dhungana2020peer}. \looseness=-1

Several studies used incentives to engage users in crowdsourcing and crowdsensing tasks \cite{capponi2019survey}\cite{muldoon2018survey}. There are a variety of incentive approaches including auction-based, reputation-based, services as incentives, social incentives, and gamification. An auction-based system is a platform where a user (auctioneer) post a task on the system then other users(bidders) bid to complete the task \cite{hou2019mobile}. The system then will select a user (called winner) from the bidders based on the object of the task requester. Reputation-based incentives are used to enhance the quality of the completed task or service \cite{hou2019mobile}. Ratings of the users based on their completed tasks motivate them to provide better quality. Services were used as incentives to motivate users to provide services in order to get them \cite{khan2019mobile}. This means the user may act as a provider or a consumer. Social incentives leverage social networks and events among users to promote global cooperation \cite{yang2017promoting}. Gamification turn tasks into playable games to attract and motivate users to provide their services. To the best of our knowledge, there is no previous work done on incentivizing users to provide wireless energy sharing\cite{dhungana2020peer}. \textit{This paper hence, the first attempt to select and compose energy services based on an incentive model.} 
\section{Conclusion}
We proposed an Incentive-based energy service requests composition framework. A new incentive model was designed that considers the attributes that affect the willingness of the provision of energy. An incentive-driven composition of Energy Service requests was proposed. The approach selects the energy service requests that maximize the reward of the provider in order to overcome the resistance of provision.  The efficiency of the proposed approach was tested against Brute  Force  (BF)  and  First  Come  First  Served(FCFS) approaches. Experimental results showed that the proposed composition outperforms the FCFS approach in maximizing the reward and outperforms the BF approach in execution time while providing very similar rewards for the provider. Future direction is to improve the framework to accommodate different incentive preferences.

\def\IEEEbibitemsep{0pt plus 1pt}

\bibliographystyle{IEEEtran}
\bibliography{ref}

\end{document}